\documentclass[longauth]{aa}
\usepackage{natbib}
\usepackage{longtable,lscape}
\usepackage{amsmath}
\bibpunct{(}{)}{;}{a}{}{,}
\usepackage{graphicx}

\usepackage[colorlinks=true,citecolor=blue]{hyperref}
\usepackage{color}
\usepackage{xspace}
\usepackage{dcolumn}
\usepackage{longtable}
\usepackage{lscape}
\usepackage{wasysym}
\usepackage{rotating}
\usepackage{afterpage}
\usepackage{mathtools}

\newcommand{\MJup}{M$_{\mathrm{Jup}}$\xspace}
\newcommand{\RJup}{R$_{\mathrm{Jup}}$\xspace}

\newcommand{\MSun}{M$_{\odot}$\xspace}

\newcommand{\mic}{$\mu$m\xspace}
\newcommand{\as}{\hbox{$^{\prime\prime}$}\xspace}

\usepackage[varg]{txfonts}
\begin{document}

\title{VLT/SPHERE exploration of the young multiplanetary system
  PDS70\thanks{Based on observation
    made with European Southern Observatory (ESO) telescopes at Paranal
    Observatory in Chile, under programs ID 095.C-0298(B), 1100.C-0481(D),
    1100.C-0481(L) and 1100.C-0481(M)}} 
\subtitle{}

\author{D. Mesa\inst{1}, M. Keppler\inst{2}, F. Cantalloube\inst{2},
L. Rodet\inst{3}, B. Charnay\inst{4}, R. Gratton\inst{1}, M. Langlois\inst{5,6},
A. Boccaletti\inst{4}, M. Bonnefoy\inst{3}, A. Vigan\inst{6},
O. Flasseur\inst{7}, J. Bae\inst{8}, M. Benisty\inst{3,9}, G. Chauvin\inst{3,9},
J. de Boer\inst{10}, S. Desidera\inst{1}, T. Henning\inst{2},
A.-M. Lagrange\inst{3}, M. Meyer\inst{11}, J. Milli\inst{12},
A. M\"uller\inst{2}, B. Pairet\inst{13}, A. Zurlo\inst{14,15,6},
S. Antoniucci\inst{16}, J.-L. Baudino\inst{17}, S. Brown Sevilla\inst{2},
E. Cascone\inst{18}, A. Cheetham\inst{19}, R.U. Claudi\inst{1},
P. Delorme\inst{3}, V. D'Orazi\inst{1}, M. Feldt\inst{2}, J. Hagelberg\inst{19},
M. Janson\inst{20}, Q. Kral\inst{4}, E. Lagadec\inst{21}, C. Lazzoni\inst{1},
R. Ligi\inst{22}, A.-L. Maire\inst{2,23}, P. Martinez\inst{21},
F. Menard\inst{3}, N. Meunier\inst{3}, C. Perrot\inst{4,24,25},
S. Petrus\inst{3}, C. Pinte\inst{26,3}, E.L. Rickman\inst{19},
S. Rochat\inst{3}, D. Rouan\inst{4}, M. Samland\inst{2,20},
J.-F. Sauvage\inst{27,6}, T. Schmidt\inst{4,28}, S. Udry\inst{19},
L. Weber\inst{19}, F. Wildi\inst{19}}

\institute{\inst{1}INAF-Osservatorio Astronomico di Padova, Vicolo dell'Osservatorio 5, Padova, Italy, 35122-I \\
  \inst{2}Max-Planck-Institut f\"ur Astronomie, K\"onigstuhl 17, 69117, Heidelberg, Germany \\
  \inst{3}Univ. Grenoble Alpes, CNRS, IPAG, 38000 Grenoble, France\\
  \inst{4}LESIA, Observatoire de Paris, PSL Research University, CNRS, Sorbonne Universit\'{e}s, UPMC Univ. Paris 06, Univ. Paris Diderot, Sorbonne, Paris Cit\'{e}, 5 Place Jules Janssen, 92195 Meudon, France \\
  \inst{5}Univ. Lyon, Univ. Lyon 1, ENS de Lyon, CNRS, CRAL UMR 5574, 69230 Saint-Genis-Laval, France \\
  \inst{6}Aix Marseille Univ., CNRS, CNES, LAM, Marseille, France \\
  \inst{7}Universit\'{e} de Lyon, UJM-Saint-Etienne, CNRS, Institut d'Optique Graduate School, Laboratoire Hubert Curien UMR 5516, F-42023, Saint-Etienne, France\\
  \inst{8}Department of Terrestrial Magnetism, Carnegie Institution for Science, 5241 Broad Branch Road, NW, Washington, DC 20015, USA \\
  \inst{9}Unidad Mixta Internacional Franco-Chilena de Astronomía (CNRS, UMI 3386), Departamento de Astronomía, Universidad de Chile, Camino El Observatorio 1515, Las Condes, Santiago, Chile \\
  \inst{10}Leiden Observatory, Leiden University, PO Box 9513, 2300 RA Leiden, The Netherlands \\
  \inst{11}Department of Astronomy, University of Michigan, 1085 S. University Ave, Ann Arbor, MI 48109-1107, USA \\
  \inst{12}European Southern Observatory (ESO), Alonso de Cordova 3107, Vitacura, Casilla 19001, Santiago, Chile \\
  \inst{13}SPGroup, ELEN/ICTEAM, UCLouvain, B-1348 Louvain-la-Neuve, Belgium \\
  \inst{14}Nucleo de Astronomia, Facultad de Ingenieria y Ciencias, Universidad Diego Portales, Av. Ejercito 441, Santiago, Chile \\
  \inst{15}Escuela de Ingenieria Industrial, Facultad de Ingenieria y Ciencias, Universidad Diego Portales, Av. Ejercito 441, Santiago, Chile \\
  \inst{16}INAF – Osservatorio Astronomico di Roma, via di Frascati 33, 00078 Monte Porzio Catone, Italy\\
  \inst{17}Department of Physics, University of Oxford, Oxford OX1 3PU, UK\\
  \inst{18}INAF - Osservatorio Astronomico di Capodimonte, Salita Moiariello 16, 80131 Napoli, Italy\\
  \inst{19}Geneva Observatory, University of Geneva, Chemin des Mailettes 51, 1290 Versoix, Switzerland\\
  \inst{20}Department of Astronomy, Stockholm University, Stockholm, Sweden\\
  \inst{21}Universit\'e C\^ote d'Azur, OCA, CNRS, Lagrange, France \\
  \inst{22}INAF – Osservatorio Astronomico di Brera, Via E. Bianchi 46, I-23807
Merate, Italy \\
  \inst{23}STAR Institute, Universit\'e de L\'iege, All\'ee du Six Ao\'ut 19c, B-4000, Li\'ege, Belgium \\
  \inst{24}Instituto de F\'isica y Astronomía, Facultad de Ciencias, Universidad
  de Valparaíso, Av. Gran Bretaña 1111, Valpara\'iso, Chile \\
  \inst{25}N\'ucleo Milenio Formaci\'on Planetaria – NPF, Universidad de
  Valpara\'iso, Av. Gran Bretana 1111, Valpara\'iso, Chile \\
  \inst{26}Monash Centre for Astrophysics (MoCA) and School of Physics and Astronomy, Monash University, Clayton Vic 3800, Australia \\
\inst{27}DOTA, ONERA, Universit\'e Paris Saclay, F-91123, Palaiseau France \\
\inst{28}Hamburger Sternwarte, Gojenbergsweg 112, D-21029 Hamburg, Germany \\
}

   \date{Received  / accepted }

\abstract
    {PDS\,70 is a young (5.4~Myr), nearby ($\sim$113~pc) star hosting a known
      transition disk with a large gap. Recent observations with SPHERE and
      NACO in the near-infrared (NIR) allowed us to detect a planetary mass
      companion, PDS\,70\,b, within the disk cavity. Moreover, observations in
      $H_{\alpha}$ with MagAO and MUSE revealed emission associated to PDS\,70\,b
      and to another new companion candidate, PDS\,70\,c, at a larger separation
      from the star. PDS\,70 is the only multiple planetary
      system at its formation stage detected so far through direct imaging.}
    {Our aim is to confirm the discovery of the second planet PDS\,70\,c
      using SPHERE at VLT, to further characterize its physical properties, and
      search for additional point sources in this young planetary system.
   }
    {We re-analyzed archival SPHERE NIR observations and obtained new data
      in Y, J, H and K spectral bands for a total of four different epochs.
      The data were reduced using the data reduction and handling pipeline and
      the SPHERE data center. We then applied custom routines
      (e.g. ANDROMEDA and PACO) to subtract the starlight. 
   }
    {We re-detect both PDS\,70\,b and c and confirm that PDS\,70\,c is
      gravitationally bound to the star. We estimate this second planet to be
      less massive than 5~\MJup and with a $T_{{\rm eff}}$ around 900~K. Also, it
      has a low gravity with $\log{g}$ between 3.0 and 3.5~dex. In addition,
      a third object has been identified at short separation ($\sim$0.12\as)
      from the star and gravitationally bound to the star. Its spectrum is
      however very blue, so that we are probably seeing stellar light
      reflected by dust and our analysis seems to demonstrate that it is a
      feature of the inner disk. We, however, cannot completely exclude the
      possibility that it is a planetary mass object enshrouded by a dust
      envelope. In this latter case, its mass should be of the order of few
      tens of $M_{\oplus}$. Moreover, we propose a possible structure for the
      planetary system based on our data that, however, cannot be stable on a
      long timescale.
   }
{}


\titlerunning{PDS\,70}
\authorrunning{Mesa et al.}
   \maketitle
%

\section{Introduction}
\label{intro}

PDS\,70 is a 5.4$\pm$1.0~Myr \citep{2018A&A...617L...2M} K7 pre-main sequence
star that is part of the Upper Centaurus-Lupus group \citep{2016MNRAS.461..794P}
at a distance of 113.43$\pm$0.52~pc \citep{gaia_dr1,gaia_dr2}. In the course
of the VLT/SPHERE SHINE survey \citep{2017sf2a.conf..331C},
\citet{2018A&A...617A..44K} discovered a planetary mass companion, PDS\,70\,b,
located in the transition disk surrounding this young star. \par

The presence of a circumstellar disk was first inferred by
\citet{2004ApJ...600..435M} due to the detection of a strong mid-infrared
excess and of a strong emission at millimeter wavelengths. The disk was first
resolved in $K_s$ band by \citet{2006A&A...458..317R} using VLT/NACO. The first
detection of a gap in the disk was obtained by \citet{2012ApJ...758L..19H}
exploiting H-band polarized data obtained through the Subaru/HiCIAO instrument.
Using the same data, \citet{2012ApJ...760..111D} estimated, assuming a distance
of 140~pc for the system, a gap size of around 65~au in which the dust is
depleted by a factor of $\sim$1000 with respect to the outer part of the disk.
They also found evidence of an inner disk with dimensions of the order of few au
detecting a weak near-IR excess in the spectral energy distribution (SED).
Finally, they estimated for the disk
a total dust mass of $\sim10^{-4}$~\MSun. PDS\,70 was also observed at
millimeter wavelengths by \citet{2015ApJ...799...43H} using the sub-millimiter
array (SMA) and, more recently, by \citet{2018ApJ...858..112L} using ALMA. The
ALMA observations were performed both at 0.87~mm continuum and at $HCO^{+}$ and
CO gas emission lines.
They allowed to define, in the dust continuum, the presence of a radial
gap between the inner and the outer disk at 15-60~au of separation, assuming a
distance of 140~pc, and to image different substructures as a bridge-like
feature and an azimuthal gap in the $HCO^{+}$ emission of the disk.
Further observations with ALMA, in the continuum and CO, at higher angular
  resolution were presented by \citet{2019A&A...625A.118K} and showed evidence
of a depletion of emission in the CO integrated intensity centered on the
separation of PDS\,70\,b, while the continuum peak is located at 74 au.
Moreover, they found, through hydrodynamical modelling of the gas kinematics,
that the presence of an additional low-mass companion further out than the
orbit of PDS 70 b may be required to account for the large gap width.\par
PDS\,70\,b was detected into the disk gap using NICI, NACO and VLT/SPHERE
data in the near-infrared (NIR) by \citet{2018A&A...617A..44K}.
Its colors are very red and through the photometry they were able to
estimate a mass of the order of 5-9~\MJup assuming for the companion the
same age of the system. The outer disk, as seen at NIR
wavelengths, has a radius of about 54~au assuming the updated value of
113.43~pc for the system. Moreover, for the first time they
were able to detect scattered light from the inner disk determining a radius
of less than 17~au, consistent with ALMA values. Exploiting new
SPHERE observations \citet{2018A&A...617L...2M} performed a characterization
of the orbital properties of PDS\,70\,b constraining its semi-major axis around
22~au corresponding to an orbital period of $\sim$118~years. Using the planet
infrared spectrum and atmospheric models, they found for the companion a
$T_{{\rm eff}}$ between 1000 and 1600~K, low surface gravity ($<$3.5~dex) and
an unusually large radius between 1.4 and 3.7~\RJup. The latter
could be a hint of the presence of circumplanetary material. Recent
observations in H and K band with VLT/SINFONI may indicate the presence of a
circumplanetary disk around PDS\,70\,b to explain its very red spectrum
\citep{2019ApJ...877L..33C,2019MNRAS.486.5819C}. \par
Following the PDS\,70\,b discovery, observations at $H\alpha$ wavelength with
the Magellan Adaptive Optics (MagAO) system \citep{2018ApJ...863L...8W} revealed
a source in $H\alpha$ at the position of the companion. This is a
hint that PDS\,70\,b is still accreting material and is consistent with the
presence of a circumplanetary disk. The authors were 
able to derive a value of $10^{-8\pm1}$~\MJup~yr$^{-1}$ for the mass accretion
rate. This result was later independently confirmed with VLT/MUSE observations
\citep{2019NatAs.tmp..329H} who redetected PDS\,70\,b in $H\alpha$ and derived
an accretion rate of $2\times10^{-8\pm0.4}$~\MJup~yr$^{-1}$. A new clear
point-source emission was detected with a signal-to-noise (S/N) of
around 8 at larger separation ($\sim$35~au) very near to the bright west ring
of the disk as seen in projection, hinting for the presence of a second
companion in the PDS\,70 system (hereinafter PDS\,70\,c). The authors
were able to estimate an accretion rate of $10^{-8\pm0.4}$~\MJup~yr$^{-1}$ for this
object. The reanalyis of the ALMA data previously used by
\citet{2018ApJ...858..112L} and by \citet{2019A&A...625A.118K} allowed
\citet{2019ApJ...879L..25I} to detect a sub-millimeter continuum emission
associated with PDS\,70\,c interpreted as originating from a dusty
circumplanetary disk. They also detected a second compact source at close
separation from PDS\,70\,b speculating that this could come from dust orbiting
in the proximity of the planet. \par
The direct imaging signal of PDS\,70\,c was marginally identified in SPHERE NIR
data in the form of a rather elongated structure but not presented in
\citet{2018A&A...617A..44K} and \citet{2018A&A...617L...2M} because of possible
contamination from the disk. Further SPHERE follow-up observations were
performed with the aim to provide a robust independent confirmation of the
existence of PDS\,70\,c, to further characterize its nature and, finally, to
search for additional point-sources in the system. \par
In this paper, we present in Section~\ref{s:obs} the data that we use for
our analysis and the data reduction procedure. In Section~\ref{s:res} we
present our results while in Section~\ref{s:dis} we discuss them. Finally, in
Section~\ref{s:conclusion} we give our conclusions.

\section{Observations and data reduction}
\label{s:obs}


\begin{table*}[!htp]
  \caption{List and main characteristics of the SPHERE observations of PDS\,70
    used for this work. During the observation of 2019-03-06, the coronagraph
    was incorrectly positioned with an offset of $\sim$20~mas in South-East
    direction.}\label{t:obs}
\centering
\begin{tabular}{ccccccccc}
\hline\hline
Date  &  Obs. mode & Coronograph & DIMM seeing & $\tau_0$ & wind speed & Field rotation & DIT & Total exposure\\
\hline
2015-05-31  & IRDIFS      & N\_ALC\_YJH\_S & 1.20\as & 1.1 ms & 4.55 m/s &$50.6^{\circ}$ &  64 s &  4096 s \\
2018-02-24  & IRDIFS\_EXT & N\_ALC\_YJH\_S & 0.40\as & 7.0 ms & 2.85 m/s &$93.4^{\circ}$ &  96 s &  6336 s \\
2019-03-06  & IRDIFS\_EXT & N\_ALC\_$K_s$  & 0.39\as & 8.9 ms & 4.45 m/s &$56.5^{\circ}$ &  96 s &  4608 s \\
2019-04-13  & IRDIFS\_EXT & N\_ALC\_YJ\_S  & 0.98\as & 2.5 ms & 8.53 m/s &$71.4^{\circ}$ &  96 s &  6144 s \\
\hline
\end{tabular}
\end{table*}

For the present work we have used both archival and new observations taken with
SPHERE \citep{2019arXiv190204080B}. The archival observations were obtained on
the nights of 2015-05-31 and 2018-02-24 and were previously used for the works
presented in \citet{2018A&A...617A..44K} and in \citet{2018A&A...617L...2M}.
In addition to these data we have also acquired new data on the nights of
2019-03-06 and 2019-04-13. The main characteristics of all these observations
are presented in Table~\ref{t:obs}. The first of these observations was carried
out in the IRDIFS mode, that is with IFS \citep{Cl08} operating in Y and J
spectral bands (between 0.95 and 1.35\mic) and IRDIS \citep{Do08} operating in
the H band with the H23 filter pair
\citep[wavelength H2=1.593~\mic; wavelength H3=1.667~\mic;][]{2010MNRAS.407...71V}. The remaining observations were performed using the IRDIFS\_EXT mode
that uses IFS in Y, J and H spectral band (between 0.95 and
1.65~\mic) and IRDIS exploiting the K band with the K12 filter pair
(K1=2.110~\mic and K2=2.251~\mic). Due to technical problems, the star
was offset by $\sim$20~mas in the South-East direction with respect to the
coronagraphic mask during the 2019-03-06 observations. In the last
observing date we used a coronagraph that was not optimized for the IRDIFS\_EXT
mode but that had a smaller inner working angle, with a diameter of 145~mas,
allowing us to observe possible objects at small separations from the central
star. This, however, did not allow to obtain good astrometric and
photometric measurements for PDS\,70\,b and PDS\,70\,c and for this reason
we were not able to use these values for this epoch in the following work.
At all the epochs, we obtained frames with satellite spots symmetric with
respect to the central star before and after the coronagraphic sequences. This
enabled us to determine the position of the star behind the
coronagraphic focal plane mask and accurately recenter the data. 
Furthermore, to be able to correctly calibrate the flux of companions, we have
acquired images with the star off-axis. In these cases, the use of appropriate
neutral density filter was mandatory to avoid saturation of the detector. \par
The data were reduced through the SPHERE data center \citep{2017sf2a.conf..347D}
applying the appropriate calibrations following the data reduction and handling
\citep[DRH; ][]{2008SPIE.7019E..39P} pipeline. In the IRDIS case, the requested
calibrations are the dark and flat-field correction and the definition of the
star center. IFS requires, besides to the dark and flat-field corrections, the
definition of the position of each spectra on the detector, the wavelength
calibration and the application of the instrumental flat. On the pre-reduced
data we then applied speckle subtraction algorithms like TLOCI
\citep{2014SPIE.9148E..0UM} and principal components analysis
\citep[PCA; ][]{2012ApJ...755L..28S} as implemented in the consortium pipeline
application called SpeCal \citep[Spectral Calibration; ][]{2018A&A...615A..92G}
and also described in \citet{zurlo2014} and in \citet{mesa2015} for the IFS.
Finally, the ANDROMEDA package \citep{2015A&A...582A..89C} and the PACO package
\citep{2018A&A...618A.138F} were also applied to all datasets.


\section{Results}
\label{s:res}

\begin{figure*}[!htp]
\centering
\includegraphics[width=1\textwidth]{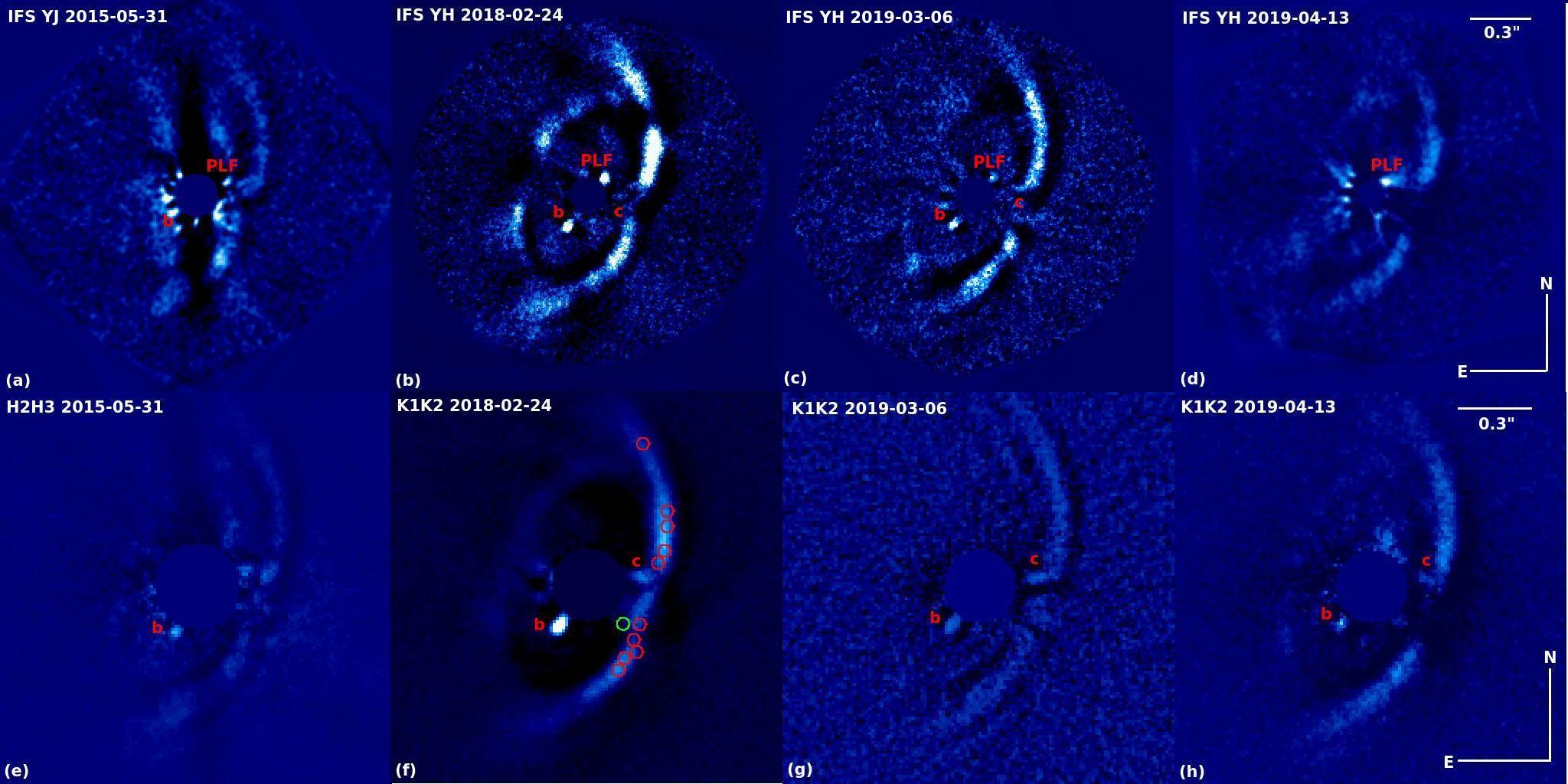}
\caption{Final IFS (upper row) and IRDIS (bottom row) images obtained for the
  following epochs: 2015-05-31 (panel {\it a} and panel {\it e}), 2018-02-24
  (panel {\it b} and panel {\it f}), 2019-03-06 (panel {\it c} and panel
  {\it g}) and 2019-04-13 (panel {\it d} and panel {\it h}). The scale and
  the orientation are the same for all the epochs and are displayed in panel
  {\it d} and in panel {\it h}. In each image we tag with red
  letters the positions of the identified companions and of the proposed point
  like feature (PLF). All the images displayed
  have been obtained using a 5 principal components PCA applied separately
  on each wavelengths apart that for the first IFS image where just 2 principal
  components were used. In panel {\it f} we overplot ten red circles to
  indicate the positions used to extract the disk spectra and one green circle
  for the position of the simulated planet as described in more details in
  Section~\ref{s:natc}.}
\label{f:finalall}
\end{figure*}

The final images, obtained using PCA are shown in Figure~\ref{f:finalall} both
for IFS (upper panels) and IRDIS (bottom panels). PDS\,70\,b is visible in
all these images but in the following we will focus on the detection of other
point sources. \par
The re-analysis of these images allowed us to detect PDS\,70\,c at a S/N ratio
of 8.6 and 9.6 at a projected separation of $\sim$0.2\as in the IRDIS data
acquired during the nights of 2018-02-24 and 2019-03-06 (panel {\it f} and
{\it g} of Figure~\ref{f:finalall}). PDS\,70\,c is visible at the same epochs
in the IFS images (panel {\it b} and {\it c} of Figure~\ref{f:finalall}) and,
barely, in the IRDIS image obtained from 2019-04-13 data (panel {\it h} of
Figure~\ref{f:finalall}). \par
A third object is clearly visible at short separation from
the star both in the 2018-02-24 and in the 2019-03-06 IFS data (panel {\it b}
and {\it c} of Figure~\ref{f:finalall}) with a S/N ratio of $\sim$5. As we
will point out in the following Sections, even if the planetary nature of this
object cannot be fully excluded, our analysis favours that this object is an
inner disk feature. For this
reason we will hereinafter refer to this detection as a point-like feature
(PLF). The source was also retrieved in the IFS 2019-04-13 observation
(panel
{\it d} of Figure~\ref{f:finalall}), during which we used a set-up optimized to
retrieve objects at very close separation from the star, as explained above.
Furthermore, we were able to find the same object at very low S/N in an
older observation (2015-05-31, see panel {\it a} of Figure~\ref{f:finalall}).
This allowed us to expand the time range of our observations. For
the IRDIS data, this object can only be barely retrieved in the 2018-02-24 data
while it is not visible in the remaining epochs. However, given the poor
detection, the photometry could not be extracted from these data either using
the negative planet method or the ANDROMEDA package. It was instead possible
to retrieve a photometry using the PACO package. \par
We have then combined the images at different epochs to obtain a complete
vision of the PDS\,70 planetary system and to determine in a more precise way
its structure following the method described in \citet{2019A&A...623A.140G}. We
have used the IFS images because it was possible to clearly image all of the
proposed companions around the star. As a first step we deprojected our data,
assuming a disk inclination and a position angle of $49.7^{\circ}$ and
$158.6^{\circ}$ respectively \citep{2012ApJ...758L..19H,2019A&A...625A.118K}, to
be able to see the disk plane face-on. We then rotated the images according to
circular Keplerian motion to have the three objects in the same positions at
different epochs and sum the images. Given the short total orbital rotation of
the companions in the considered time range, the effects of the uncertainties on
the stellar and companion masses are not relevant in the final result of this
procedure. Finally, we projected back our data to obtain a view of the system as
seen on-sky. \par
\begin{figure*}[!htp]
\centering
\includegraphics[width=1\textwidth]{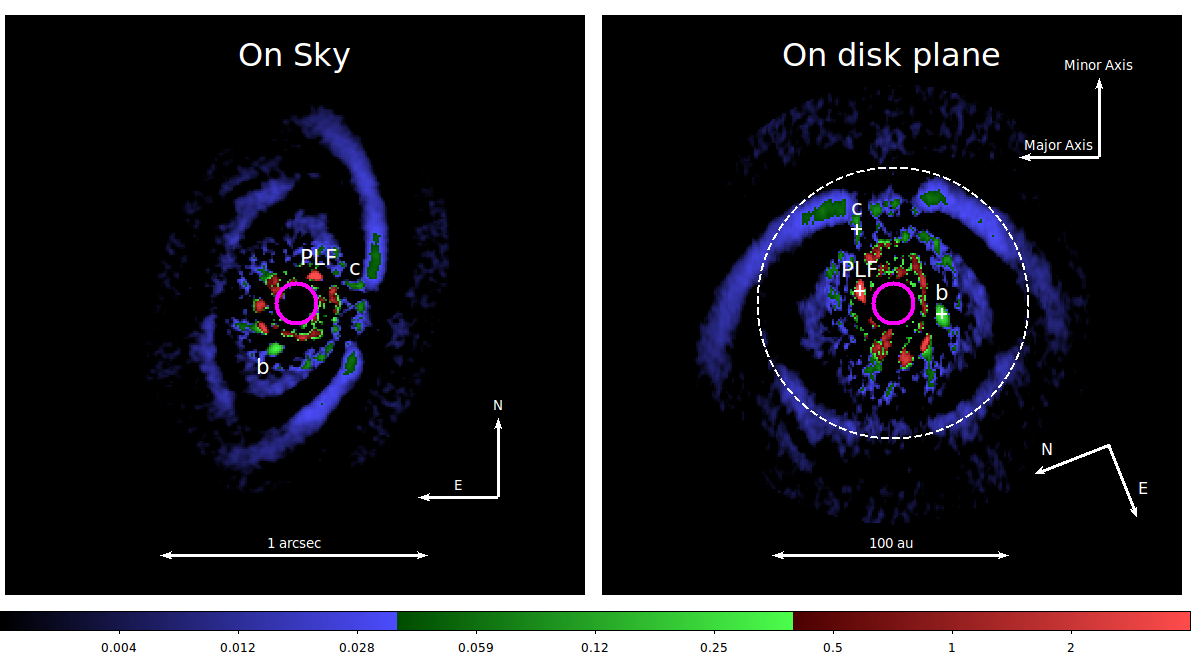}
\caption{{\it Left:} IFS image obtained combining the images from all the
  epochs as described in the text. The positions of the three objects known
  around PDS\,70 are tagged. {\it Right:} Deprojected image of the PDS\,70
  system as seen on the disk plane. The dashed circle indicate the position of
  the inner edge of the outer ring of the disk as found with ALMA by
  \citet{2019A&A...625A.118K} to show that it corresponds to the partially
  imaged ring in the SPHERE images.}
\label{f:comboimg}
\end{figure*}

The final result of this procedure is shown in
Figure~\ref{f:comboimg} where in the left panel we show the on-sky view of the
system with the indication of the positions of the three proposed objects.
On the right panel, we instead show the deprojected view of the system seen in
the disk plane. We also plotted a dashed circle corresponding to the expected
position of the inner edge of the outer disk inferred from the sub-millimeter
continuum observations \citep{2019A&A...625A.118K} to show
that it corresponds to the eastern part of the ring visible in the SPHERE image.
Furthermore, we note that the position of the western outer ring of the disk,
apparently very near to the position of PDS\,70\,c is due to a projection
effect because the bright western part of the disk seen in scattered light is
actually the upper layer of a torus at certain height with respect to the disk
plane. This is similar to other cases observed with SPHERE like e.g.
RX\,J1615.3-3255 \citep{2016A&A...595A.114D} and HD\,100546
\citep{2018A&A...619A.160S}. \par
A complete discussion of the structure of the disk is however beyond the
scope of this paper, which is focused on the confirmation and characterization
of PDS\,70\,c and on the additional point-like feature in this system.



\subsection{PDS\,70\,b}

The new SPHERE data did not allow us to update the spectral results given in
\citet{2018A&A...617L...2M} for PDS\,70\,b. We then refer to this paper
for any information about the spectral analysis for this object. \par
We used our recent SPHERE data to update the orbital solutions of PDS\,70\,b
which are in good agreement with what was found by \citet{2018A&A...617L...2M}.
The results of this analysis are given in the top right panel of
Figure~\ref{f:astroplanall}, in Section~\ref{s:struct} and in
Section~\ref{s:stability}.

\subsection{PDS\,70\,c}
\label{s:companionc}

\begin{figure*}[!htp]
\centering
\includegraphics[width=1.\textwidth]{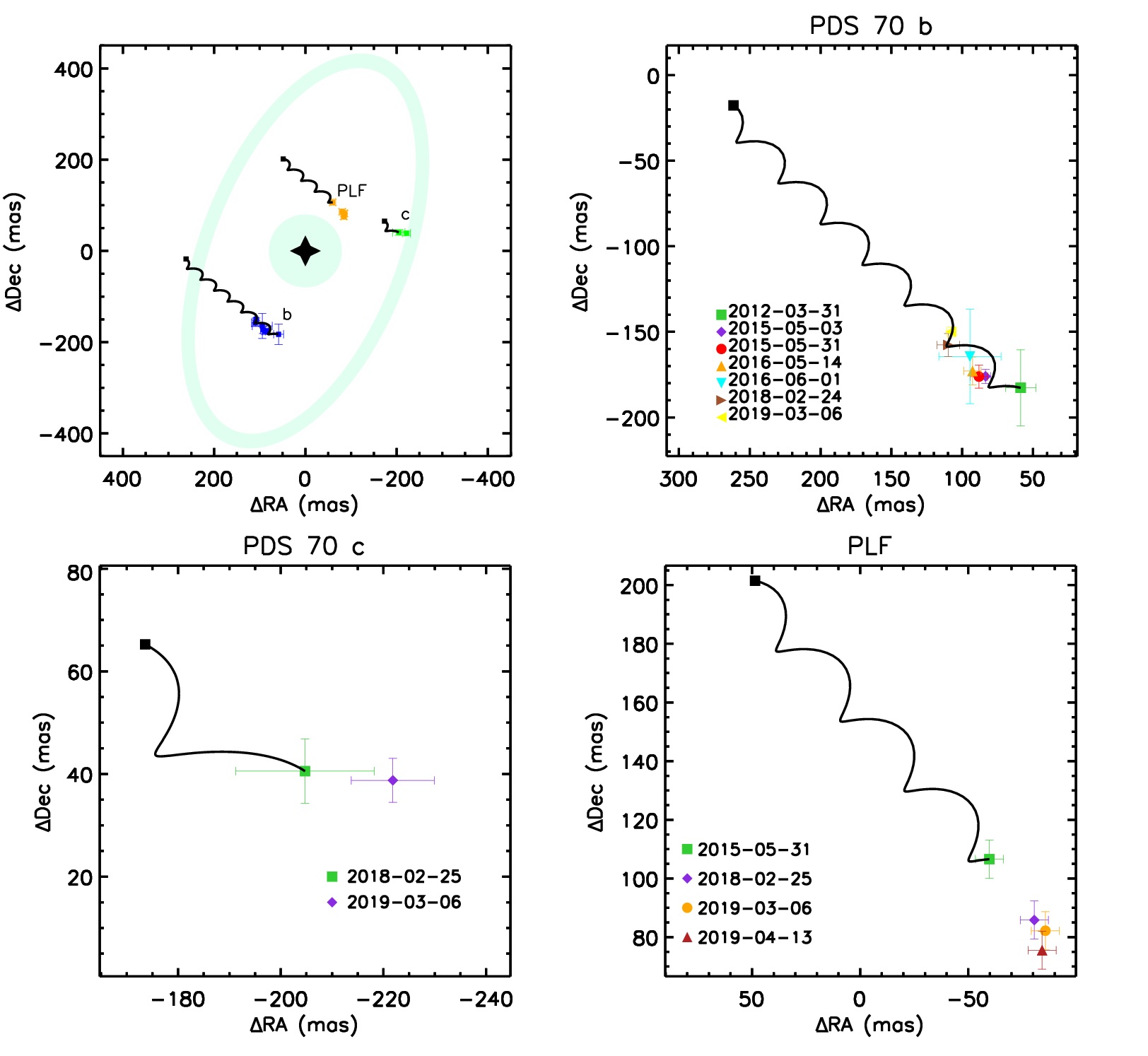}
\caption{{\it Top left panel: } Relative astrometric positions of the three
  proposed companions of PDS\,70 with respect to the host star, represented by
  the black star symbol. The positions of the inner and of the outer disk are
  also displayed through light cyan areas. The solid black lines represent the
  expected course of the companion if it were a background object. The length
  differences between the black lines for different objects are due
  to the different temporal coverage that we have for each of them. The black
  squares at the end of the line represent the expected position at the epoch of
  the last observation in this case. For clarity, we represent a zoomed
  version of the areas around each companion in the {\it Top right panel} for
  PDS\,70\,b, in the {\it Bottom left panel} for PDS\,70\,c and in the
  {\it Bottom right panel} for the PLF.}
\label{f:astroplanall}
\end{figure*}

The main limitation to the analysis of this object with SPHERE is that it is
projected very near to the bright western ring of the disk. This
prevented a robust identification in previous analysis as it was regarded as
a feature of the disk and, moreover, complicating both the astrometric and
photometric extraction. In our new analysis, we were able
to obtain these values using the negative planet method
\citep[see e.g., ][]{2011A&A...528L..15B,zurlo2014} for the two best
observing epochs (2018-02-24 and 2019-03-06) and confirming that it is
actually a planet. In Table~\ref{t:astroc} we list the astrometric values
obtained from the IRDIS data for the two available epochs through this
approach. The relative positions of PDS\,70\,c are also shown in the bottom
left panel of Figure~\ref{f:astroplanall} where they are compared to the
positions expected for a background object. This allows us to exclude that
this is a background object. \par
\begin{figure}
\centering
\includegraphics[width=1.\columnwidth]{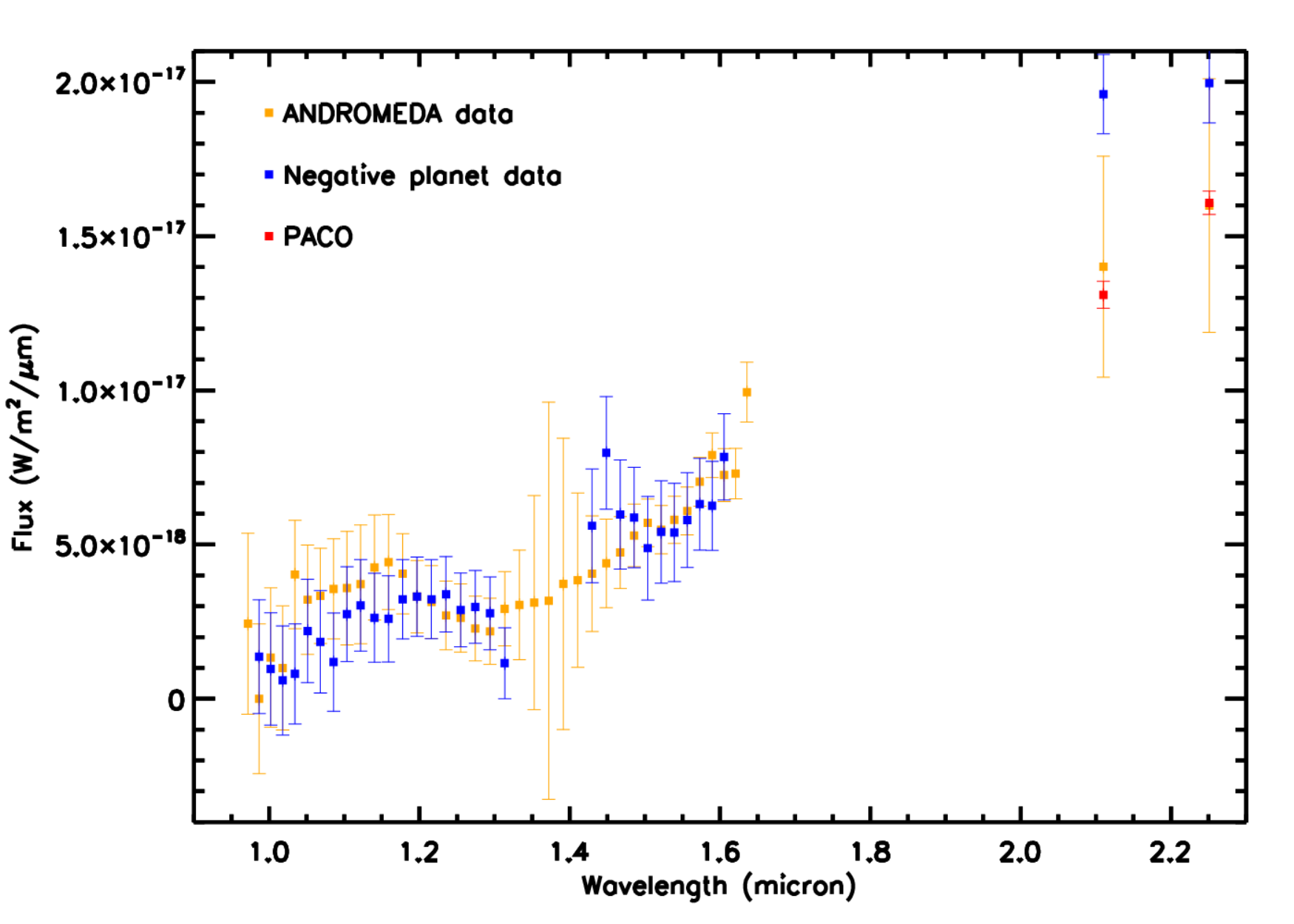}
\caption{Comparison between the PDS\,70\,c spectrum extracted using the
  negative planet method (blue squares), ANDROMEDA (orange squares) and PACO
  (magenta squares). For the latter method, we plot only the two values
  obtained from IRDIS data. Due to fact that in the region around 1.4~\mic the
  negative planet method gives negative values, those points are not shown in
  the plot.}
\label{f:comparemethod}
\end{figure}
\begin{figure*}
\centering
\includegraphics[width=1\textwidth]{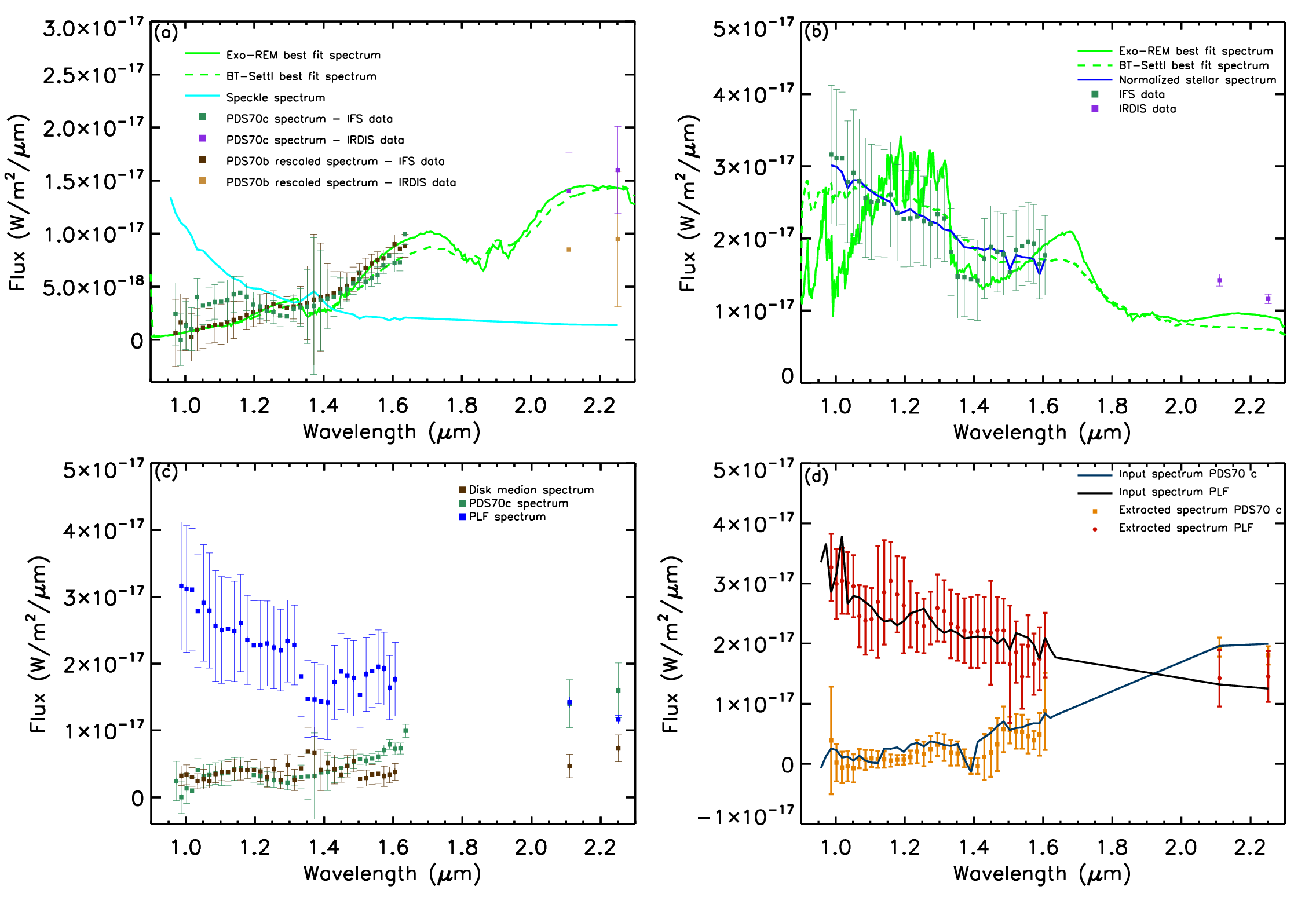}
\caption{{\it Panel (a):} Extracted spectrum of PDS\,70\,c both with
  IFS (green squares) and IRDIS (violet squares). The green solid line
  represents the best fit Exo-REM model while the green dashed line is the
  best fit BT-Settl model. The parameters of the best-fit models are listed
  in the text (Section~\ref{s:natc}). The cyan solid line represents an
  extracted spectrum for a speckle from the same dataset. We also overplot the
  rescaled spectrum of PDS\,70\,b both for IFS (deep brown squares) and
  IRDIS (light brown squares). {\it Panel (b): } Same
  as {\it panel (a)} but for the PLF spectrum. We plot as a blue solid line the
  stellar spectrum normalized to the companion flux to demonstrate their
  similarity.
  {\it Panel (c): } Comparison between the extracted spectra for PDS\,70\,c
  and the PLF and the spectrum obtained making a median of ten different
  positions of the disk. {\it Panel (d):} Comparison between the injected
  spectrum of a simulated planet in a position similar to PDS\,70\,c with
  respect to the disk (deep blue solid line) and the relative extracted
  spectrum (orange squares). The same thing is done for a simulated planet
  at a similar position of the PLF (black solid line and red circles).}
\label{f:spectrumall}
\end{figure*}

The negative planet method was also used to calculate the photometric values.
However, due to the faintness of this object in the Y, J and H spectral bands,
we had to perform this method using the average over 5 consecutive spectral
channels. This resulted in the fact that we could not derive any photometric
value for the first two and the last two IFS channels. The final spectrum is
obtained through a weighted average, with the weight considering the different
qualities of the observations, of the results obtained for the last
three epochs considered here. To check the reliability of this result we have
extracted the spectrum also using the ANDROMEDA and the PACO package.
ANDROMEDA gave results comparable to those from the first method as shown in
Figure~\ref{f:comparemethod}, apart from the fact that the negative planet
method gives negative values in the spectral region around 1.4~\mic. These
values are likely related to water
telluric absorption affecting both the stellar and the planet flux but
we cannot exclude that they are linked to the effect of the data reduction
method. The same feature is not present in the ANDROMEDA spectrum. On the other
hand, the spectrum extracted with PACO had a detection limit below 3 sigma
so that we did not consider it as a reliable detection and we do not use it
for this work. Instead,
its results for the IRDIS data are in good agreement with those for ANDROMEDA as
shown in Figure~\ref{f:comparemethod}. For all the reason listed above, we will
therefore use in the following just the results from the ANDROMEDA package.
The result of this procedure is displayed in panel {\it a} of 
Figure~\ref{f:spectrumall} where the green squares are the photometric values
obtained from IFS while the violet squares are obtained from IRDIS. The
extracted spectrum displays very red colors providing further confirmation of
the planetary nature of this object as also confirmed by comparing it to the
solid cyan line that represents the extracted spectrum for a speckle from the
same dataset, obtained before applying the speckle subtraction procedure at a
separation comparable to that of the companion. Using these results we can
also derive the photometric results in different spectral bands that are listed
for the two epochs in which it is visible in Table~\ref{t:astroc}. \par

\begin{table*}[!htp]
  \caption{Astrometric and photometric (absolute magnitudes) results obtained
    for PDS\,70\,c.}
  \label{t:astroc}
\centering
\begin{tabular}{ccccccccc}
\hline\hline
Date  &  $\Delta$RA (\as)  & $\Delta$Dec (\as)  & $\rho$ (\as) & PA & $M_J$  & $M_H$  &  $M_{K1}$  &   $M_{K2}$  \\
\hline
2018-02-24  & -0.205$\pm$0.013 & 0.041$\pm$0.006 & 0.209$\pm$0.013 & 281.2$\pm$0.5 & 17.45$\pm$0.33  & 14.89$\pm$0.87 &  12.69$\pm$0.17  &  12.53$\pm$0.19 \\
2019-03-06  & -0.222$\pm$0.008 & 0.039$\pm$0.004 & 0.225$\pm$0.008 & 279.9$\pm$0.5 & 17.00$\pm$0.40  & 14.90$\pm$0.90 &  12.49$\pm$0.11  &  12.25$\pm$0.14 \\
\hline
\end{tabular}
\end{table*}

\subsection{PLF}
\label{s:companiond}

Like for the case of PDS\,70\,c we extracted astrometric and photometric
values for the PLF following the same methods. In this case however, we were
able, using only the IFS data, to obtain the astrometric values in all
four observing epochs exploited for this work. This supports the fact that the
signal is not an artefact from the data processing. The astrometric values for
this object are listed in Table~\ref{t:astrod} while its relative positions are
compared to those expected for a background object in the bottom right panel
of Figure~\ref{f:astroplanall}. Also in this case the analysis confirms that
this source is gravitationally bound to PDS\,70. As done for PDS\,70\,c,
the spectrum extraction was performed both using the negative planet procedure,
the ANDROMEDA package and the PACO package leading to comparable results for
the IFS data. Due to the fact that it was possible to detect this
object only marginally in IRDIS data, it was not possible to extract any
reliable photometry with the first two methods while this was possible using
the PACO package. We then use these values for the spectrum of this
object that is shown in panel {\it b} of Figure~\ref{f:spectrumall}. PACO
could lead to an underestimation of the error bars with respect to other
methods due to a different way to calculate them as can be seen in
Figure~\ref{f:spectrumall}. While this can lead to some inconsistency, we
included its results to give a more complete view of the characteristics of
this object. It appears
blue and very similar to the stellar spectrum at these wavelengths as
demonstrated by the blue solid line plotted in the same Figure obtained using,
following \citet{2018A&A...617L...2M}, a calibrated spectrum of PDS\,70 from
the SpeX spectrograph. To be able to overplot the two
spectra, we have normalized the stellar spectrum to the flux of the companion.
This is a strong hint that the PLF spectrum is probably mainly due to
the reflection of the stellar light from dust.

\begin{table*}[!htp]
  \caption{Astrometric results obtained for the PLF.}\label{t:astrod}
\centering
\begin{tabular}{ccccc}
\hline\hline
Date  &  $\Delta$RA (\as)  & $\Delta$Dec (\as)  & $\rho$ (\as) & PA \\
\hline
2015-05-31  & -0.060$\pm$0.007 & 0.107$\pm$0.007 & 0.122$\pm$0.007 & 330.7$\pm$0.5 \\
2018-02-24  & -0.081$\pm$0.004 & 0.086$\pm$0.004 & 0.118$\pm$0.004 & 316.8$\pm$0.5 \\
2019-03-06  & -0.086$\pm$0.004 & 0.082$\pm$0.004 & 0.119$\pm$0.004 & 313.8$\pm$0.5 \\
2019-04-13  & -0.084$\pm$0.004 & 0.076$\pm$0.004 & 0.113$\pm$0.004 & 311.7$\pm$0.5 \\
\hline
\end{tabular}
\end{table*}



\subsection{Structure of the PDS\,70 planetary system}
\label{s:struct}

To constrain the fundamental orbital parameters for the three candidate
companions, we run a Monte-Carlo simulation using the Thiele-Innes
formalism \citep{1960pdss.book.....B} as devised in
\citet{2011A&A...533A..90D}, \citet{2013A&A...554A..21Z} and
\citet{2018MNRAS.480...35Z} and adopting the convention by
\citet{2000eaa..bookE2855H}. The simulation generates $5\times10^7$ orbits and
rejects all those that do not fit the astrometric data. With this aim, we
assumed for the star a mass of 0.76~\MSun and a distance of 113.43~pc.
Moreover, we assumed a circular orbit for each companion. In this way, we
found for PDS\,70\,b an orbital radius of $22.7^{+2.0}_{-0.5}$~au, a period of
$123.5^{+9.8}_{-4.9}$~years and an orbit inclination of $39.7^{\circ+5.4}_{-2.8}$, in
good agreement with what was found by \citet{2018A&A...617L...2M}. Due to the
paucity of astrometric points, the results for PDS\,70\,c are less defined than
for PDS\,70\,b. We obtained a radius of $30.2^{+2.0}_{-2.4}$~au, a period of
$191.5^{+15.8}_{-31.5}$~years and an inclination of
$40.8^{\circ+30.3}_{-14.0}$~degrees.
Finally, we obtained for the PLF orbit a radius of $13.5^{+0.3}_{-0.2}$~au, a
period of $56.3^{+2.0}_{-1.1}$ years and an inclination of
$46.5^{\circ+2.3}_{-10.4}$~degrees. \par
We combined these results together with those for the disk
\citep{2019A&A...625A.118K} to generate a model for the structure of the
planetary system around PDS\,70. The values of the estimated separation
expressed in au both for the companion and for the part of the disk, together
with the Keplerian rotation period, the companion masses as calculated in
Section~\ref{s:natc} and \ref{s:natd} and with the ratio
between the object period and the period of PDS\,70\,b are listed in
Table~\ref{t:struct}. From these data we can see that, if the PLF is a planet,
its orbit is in 2:1 resonance with the outer edge of the inner disk while a 3:2
resonance can be found between the orbits of PDS\,70\,b and PDS\,70\,c.

\begin{table*}[!htp]
  \caption{Structure of the PDS\,70 planetary system. In second columns we list
    the masses of the of the companions including also that of the PLF as it was
    point source. In third column we list the separations both for companions
    and for inner and outer disk. In fourth column we listed the most probable
    periods (for the disk part we assume a keplerian motion). Finally, in fifth
    column we list the ratio between the period of the companion/disk and the
    period of PDS\,70\,b to evidence possible orbital resonance.
  }\label{t:struct}
\centering
\begin{tabular}{ccccc}
\hline\hline
Object  &  Mass (\MJup)  & Separation (au) & Period (year)  & $P_{obj}$/$P_{PDS\,70\,b}$ \\
\hline
Inner disk         & //        &  9.1 &  31.5 &  0.26 \\
PLF                & 0.05-0.28 &  $13.5^{+0.3}_{-0.2}$ & $56.3^{+2.0}_{-1.1}$  &  0.46 \\
PDS\,70\,b         & 5.0-9.0   &  $22.7^{+2.0}_{-0.5}$ & $123.5^{+9.8}_{-4.9}$ &  1.00 \\
PDS\,70\,c         & 4.4$\pm$1.1& $30.2^{+2.0}_{-2.4}$ & $191.5^{+15.8}_{-31.5}$ &  1.55 \\
Outer disk (ring 1)& //        & 60.0 & 533.1 &  4.32 \\  
Outer disk (ring 2)& //        & 74.0 & 730.2 &  5.91 \\
\hline
\end{tabular}
\end{table*}


\section{Discussion}
\label{s:dis}

\subsection{On the nature of PDS\,70\,c}
\label{s:natc}

The projected position of PDS\,70\,c is very close to the bright West
part of the disk. This rises the concern that the spectrum shown in the
panel {\it a} of Figure~\ref{f:spectrumall} can be contaminated and hence
biased by the disk signal. To check this possibility we
have extracted the spectrum of the disk (using both IFS and IRDIS photometry)
at ten different positions of the disk as indicated by the red circles in
panel~{\it f} of Figure~\ref{f:finalall}. Due to the fact that we are extracting
the photometry of an extended structure, it is not possible to use the negative
planet procedure as we did for PDS\,70\,c. We then applied an aperture
photometry method, using an aperture with a radius of three pixels.
We then made a median of these spectra to compare the resulting spectrum
with that extracted for the companion. The result of this
procedure is displayed in panel {\it c} of Figure~\ref{f:spectrumall} where
the spectra of the companion and of the disk are represented
by the green and the orange line, respectively. We can see that, while at
short wavelengths the two spectra are very similar, they diverge starting from
the IFS H-band and are very different in the IRDIS K-band. As a further test,
to check if the disk can influence the companion spectrum extraction, we have
injected in our datacube a simulated planet in a position similar to PDS\,70\,c
with respect to the disk. The position of the simulated companion is indicated
by a green circle in panel {\it f} of Figure~\ref{f:finalall}. The input
spectrum of the simulated planet was
obtained scaling the spectrum of PDS\,70\,b. After that, we extracted the
companion spectrum using the same method as used for PDS\,70\,c. The result of
this procedure is displayed in panel {\it d} of Figure~\ref{f:spectrumall}
where the input spectrum is represented by the blue line while the orange
squares are the extracted photometric values. From these results we can conclude
that we are able to faithfully reproduce the input spectrum of the simulated
planet. Given the similarity of this case with that of PDS\,70\,c, we conclude
that the results of these tests support the hypothesis that the signal comes
from a point-like source not related to the circumstellar disk and is of
planetary origin. However, it is important to stress that in the Y- and
J-spectral bands, the signals from the companion and from the disk are both
very low and virtually indistinguishable so that, while we will use them to
define the physical characteristics of the companion, the final results should
be regarded with caution providing just upper limits. \par
Moreover, we can check the reliability of the extracted spectrum by comparing it
to the spectrum of PDS\,70\,b. In panel {\it a} of Figure~\ref{f:spectrumall}
we overplot the extracted spectrum for PDS\,70\,b, rescaled to be able to
compare it with that of PDS\,70\,c. For this purpose, we used a calculated
rescaling factor of 0.18. The two spectra are very similar indicating that both
these objects possess red and dusty atmospheres and/or circumplanetary disks.


We have estimated the mass of PDS\,70\,c by comparing the photometric values
reported in Table~\ref{t:astroc} with the AMES-DUSTY models
\citep{2001ApJ...556..357A} assuming the system age of 5.4$\pm$1.0~Myr. The
error bars on the age could be underestimated for late-type Scorpius-Centaurus
stars due to the fact that the effects of the magnetic activity are not taken
properly into account as found by \citet{2019A&A...622A..42A} for the case of
HIP\,79124. The results of
this procedure are listed in Table~\ref{t:massc}. The error bars are obtained
taking into account the uncertainties both on the distance of the system, on
the age of the system and on the extracted photometry. The mass values
estimated are in good agreement between the different wavelength bands and
point toward a planetary object with a mass just above 4~\MJup. We notice
here that these determinations assume that the observed magnitudes are due to
the planetary atmosphere and that the age of the companion is the same as
the age of the star. 
\par
\begin{figure*}
\centering
\includegraphics[width=0.9\columnwidth]{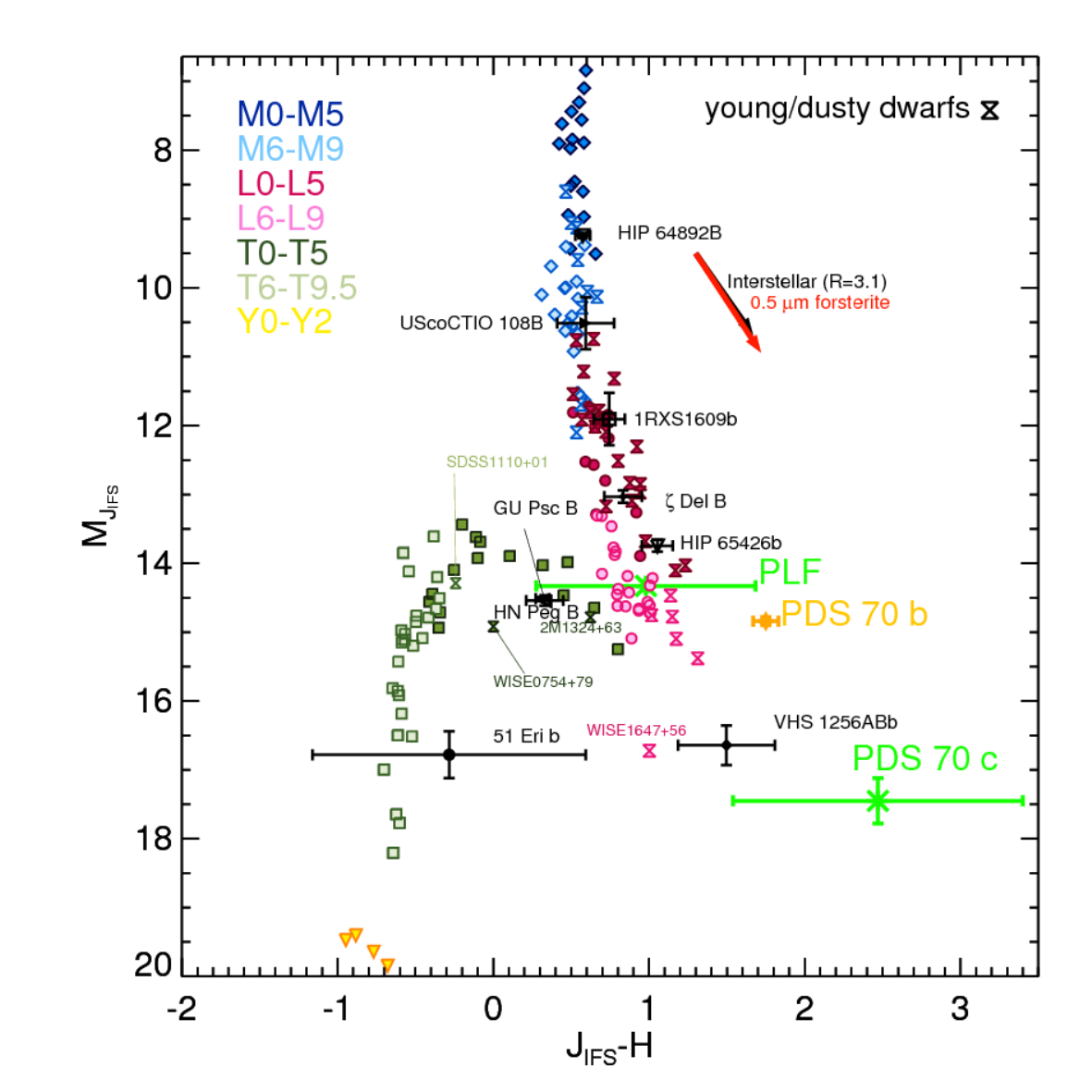} 
\includegraphics[width=0.9\columnwidth]{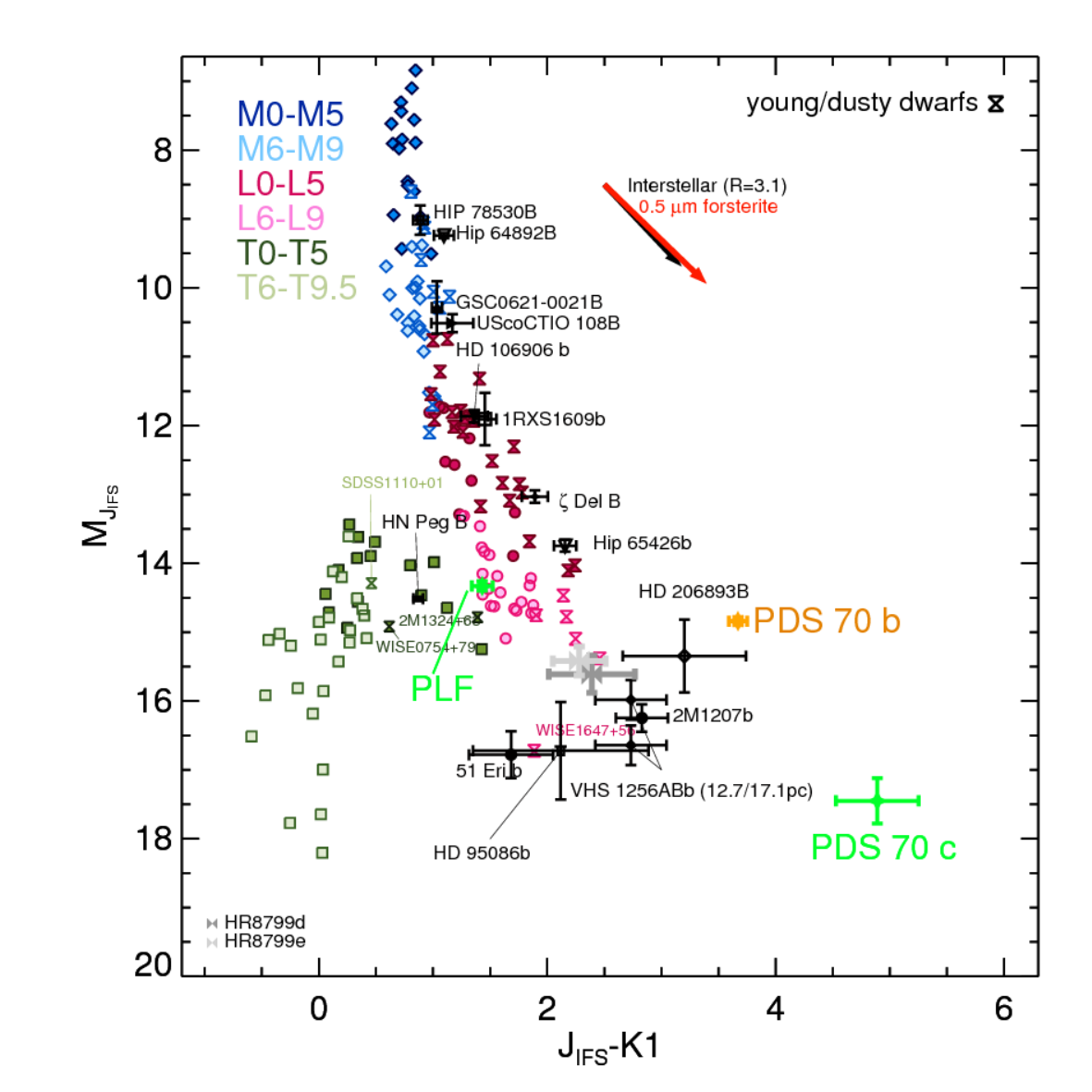}
\caption{{\it Left:} Positions of PDS\,70\,c and of the PLF (green stars)
  in the J-H versus J color-magnitude diagram. The position of PDS\,70\,b
  (orange star) is also indicated as comparison. We also display the positions
  of field dwarfs indicated with different symbols according to the spectral
  type and of some low mass companions. Moreover, we overplot the reddening
  vector computed for the synthetic extinction curve using $R_V$=3.1 (black
  arrow) and the reddening vector due to the extinction curve of forsterite
  (red arrow). {\it Right: } Same as left
  panel for the J-K1 versus J diagram.}
\label{f:cmdc}
\end{figure*}

We have also used the extracted photometry to build two 
color-magnitude diagrams. They were created following the procedure described
in \citet{2018A&A...618A..63B} and they are displayed in the left (J-H versus
J) and in the right (J-K1 versus J) panels of Figure~\ref{f:cmdc}. The
positions of PDS\,70\,c in these two diagrams demonstrate that it is an
extremely red object even with respect to PDS\,70\,b that is also shown for
comparison. The larger redness of PDS\,70\,c is also confirmed by the
comparison of the spectra of these two objects in panel {\it a} of
Figure~\ref{f:spectrumall} where the K-band values for PDS\,70\,b are lower
than those for PDS\,70\,c when the rescaling factor is taken into account. 
We overplot on both color-magnitude diagrams the reddening vectors due to
interstellar extinction \citep{2003ARA&A..41..241D} and by 0.5~\mic forsterite
grains using the optical constants by \citet{1996ApJS..105..401S}. These grains
are proposed to explain the red colors of dusty and variable L dwarfs
\citep{2014MNRAS.439..372M,2016A&A...587A..58B}. Both PDS\,70\,b and PDS\,70\,c
appear to be shifted along these vectors with respect to the sequence of
field dwarfs. These results are compatible with both these objects being
low mass objects strongly reddened by the presence of dust.

\begin{table}
  \caption{Mass estimation for PDS\,70\,c obtained from the SPHERE photometry
  and from the AMES-DUSTY models. }\label{t:massc}
\centering
\begin{tabular}{cc}
\hline\hline
Spectral band  &  Mass (\MJup)  \\
\hline
  J  &  4.35$\pm$0.45  \\
  H  &  4.21$\pm$0.78  \\
  K1 &  4.31$\pm$0.45  \\
  K2 &  4.02$\pm$0.41  \\
\hline
\end{tabular}
\end{table}

With the aim to characterize this companion we fitted its extracted spectrum
with Exo-REM models taking into account the presence of thick clouds
\citep{2015A&A...582A..83B,2017sf2a.conf..343B,2018ApJ...854..172C}. The
atmospheric models were calculated on a grid
with $T_{{\rm eff}}$ varying between 300 and 2000~K with steps of 50~K and with
$\log{g}$ with values between 3.0 and 6.0~dex with steps of 0.1~dex. On the
other hand, we verified that the metallicity did not influence the final
result so that we assumed solar metallicity (we explored also the 0.3x and 3x
solar metallicity). As can be seen in the panel {\it a}
of Figure~\ref{f:spectrumall}, where the orange solid line represents the best
fit Exo-REM model, the fit is good along all the extracted spectrum.  Best
fits are found for models with low gravity ($\log{g}\sim3.0-3.5$) and for
$T_{{\rm eff}}$ in the range 800-1100~K. The masses range between 0.5 and 4~\MJup
and are just partially in agreement with what we found previously with
AMES-DUSTY models. The best fit model represented in Figure~\ref{f:spectrumall}
has $T_{{\rm eff}}$=900~K, $\log{g}$=3.1~dex, R=1.95~\RJup and M=1.93~\MJup. \par
To confirm these results we performed a similar procedure using the BT-Settl
models \citep{2014IAUS..299..271A} with $T_{{\rm eff}}$ varying between 700 and
2500~K
with a step of 100~K and a $\log{g}$ between 3.0 and 5.5~dex with a step of
0.5~dex. Also in this case we considered only solar metallicity. We obtained
good fit for $T_{{\rm eff}}$ between 800 and 1100~K and for $\log{g}$=3.5.
We can then conclude that the two models are in good agreement as also
confirmed by the best fit model also displayed in Figure~\ref{f:spectrumall}
with the green solid line. In this case the best fit is obtained for
$T_{{\rm eff}}$=900~K and $\log{g}$=3.5~dex. \par
The recently suggested presence of a circumplanetary disk
\citep{2019ApJ...879L..25I} may cause selective absorption, reddening the
object spectrum. This would explain, at least partially, the very red colors
that we have found in our work and it would confirm recent results on the
detectability of circumplanetary disks from hydrodynamic simulation
\citep{2019arXiv190601416S,2019MNRAS.487.1248S}. The effects of the
circumplanetary disk should be carefully modeled because they could partially
modify the fitting results that we have found in this Section but this analysis
is beyond the scope of the present work. The results presented here should be
regarded as a first estimation of the companion physical characteristics.


\subsection{On the nature of the PLF}
\label{s:natd}

The fact that the PLF is recoverable in at least four different epochs is
a clear evidence that it is a real astrophysical signal. Moreover, its
astrometric positions in the four epochs clearly exclude the possibility that
it is a background object and they are in agreement with a Keplerian circular
motion around the star. However, as discussed above, its spectrum is blue and
very similar to the stellar spectrum. To confirm the reliability of the
extracted spectrum we have extracted the spectrum of a simulated planet
with a spectrum similar to that of the PLF and in a position similar to that of the PLF following the same method used for PDS\,70\,c and described in
Section~\ref{s:natc}. The results are displayed in panel {\it d} of
Figure~\ref{f:spectrumall} where the input spectrum is represented with a
black solid line while the extracted spectrum is represented by red
circles. Like for the case of PDS\,70\,c, we are able to faithfully reproduce
the input spectrum. A blue spectrum is a strong hint that the emission
is due to stellar light reflected likely by dust. Its measured separation of
$\sim$13.5~au is not consistent with the radius of the inner disk of about 9~au
\citep{2019A&A...625A.118K} corresponding to an angular radius of
$\sim$0.08\as that should be completely behind the SPHERE coronagraph. We have
however to consider that the inner disk size is obtained in the
sub-millimeter continuum that probes the distribution of large
dust grains. We cannot exclude that the distribution of small dust grains,
probed by scattered light, could be different. \par
\begin{figure}
\centering
\includegraphics[width=0.8\columnwidth]{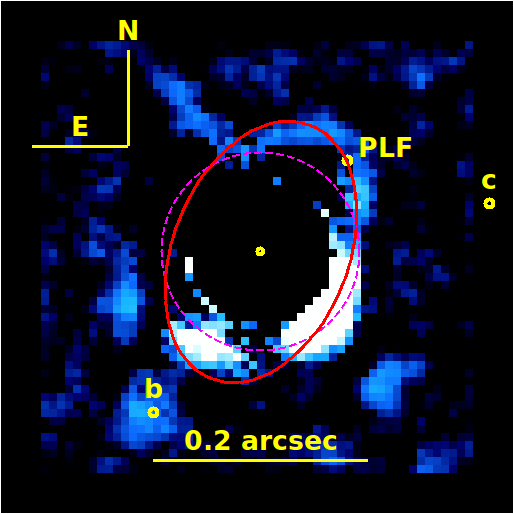}
\caption{Image resulting from the off-axis PSF subtraction described in the
  text. The red ellipse represent the estimated position for the inner disk.
  The dashed magenta circle represents the position of the coronagraph. The
  yellow circles are used to tag the positions of the point sources.}
\label{f:forward}
\end{figure}

A hint favoring this latter view comes from the analysis of the IFS off-axis
PSF images obtained just before and just after the coronagraphic observations
as explained in Section~\ref{s:obs}. Exploiting the field rotation between
these two images we can perform a subtraction to eliminate static speckles
on these images. This procedure is performed separately in Y, J and H band.
In normal conditions, the limiting contrast that we can obtain in the region
between 60 and 110~mas from the star is of the order of 6-7 magnitudes.
However, the very good conditions obtained during the night of 2018-02-24
allowed us to reach a contrast of the order of 10 magnitudes at the same
separations. In Figure~\ref{f:forward} we present the results of this procedure
on the H-band data. In this image there is a possible detection of the inner
ring of the disk. We draw an ellipse (red solid line) adopting the inclination
and the PA of 49.7$^{\circ}$ and 158.6$^{\circ}$, respectively. Moreover, its
semi-major axis is of 128 mas. This ellipse is roughly passing through the
position of the PLF. While this is a differential image, it is not an ADI image
because it uses only two frames. For this reason, we expect that the region on
the near side of the disk around the semi-minor axis (corresponding to the
forward scattering of the disk) is the brightest, as indeed it is. The contrast
value in the position of the PLF is about 9.5 mag,  consistent with the
photometry obtained on the coronagraphic image. These results therefore
appear consistent with scattered light from the inner disk. In any case, it is
important to stress that while these results are suggestive of the presence
of the inner disk, they cannot be considered as a clear detection due to the
very low SNR of the signal and to the presence of other faint structures in
the FOV. However, we would like to note that this structure has the
orientation expected for the inner disk and that a comparison with other
observations taken in similar optimal conditions as those for PDS\,70 does not
reveal similar structures. If the interpretation of this image as the inner
disk is correct, then the PLF is a part of the disk itself.

The positions of the PLF in both the color-magnitude diagrams
displayed in Figure~\ref{f:cmdc} are compatible with those
of field dwarfs with spectral type L6-L9. Moreover, we followed the same
procedure used for PDS\,70\,c fitting its extracted spectrum both with
Exo-REM and BT-Settl atmospheric models. The best fit spectra from this
procedure are shown in the panel {\it b} of Figure~\ref{f:spectrumall}.
Exo-REM models point toward a small star with a mass of $\sim$100~\MJup
and $T_{{\rm eff}}$ between 1400 and 2000~K. The value of $\log{g}$ should be of
the order of 6. On the other hand, BT-Settl models favor a much higher
$T_{{\rm eff}}$ of the order of 3000~K while the surface gravity is unconstrained.
These results are generally quite inconsistent with each other. Moreover, a
small star at such a short separation from the star would probably have a strong
effect on the stability of the system generating a strong radial velocity
signal that has never been observed. Furthermore, a stellar object would be much
brighter, of the order of 6-9 magnitudes,  than what we found for the PLF.
These considerations seem to confirm our previous conclusions that we are not
actually looking to the photosphere of a sub-stellar object but rather to
stellar light reflected by dust. If so, the inconsistent results obtained when
attempting to fit this spectrum with the atmospheric models are naturally
explained. \par
The results above seem to demonstrate that the PLF is actually a part
of the inner disk, which is supported by the fact that no $H_{\alpha}$ emission
associated to the PLF has been detected so far. Still, we cannot completely
exclude the possibility
that it is actually a point source that should be however embedded into the
disk. If this were the case, we can consider two alternative possibilities.
The first one is that it is a transient blob of dust that is destined to break
apart in a timescale of few thousands of years similar to blobs B and C seen in
the disk of HD\,169142 \citep{2018MNRAS.473.1774L,2019A&A...623A.140G} and, for
this reason, the probability to observe such a transient phenomenon is
generally low unless these structures are generated frequently. The second
possibility is that we are observing a dust envelope around a forming
planet. This case would be very similar to what was suggested for blob D around
HD\,169142 with SPHERE data \citep{2019A&A...623A.140G}. Given the spectrum, we
should not see any photospheric emission from the planet due to the fact that
the envelope is optically thick or the light reflected by the dust dominates
over the photospheric emission. If this is the case, we can estimate a guess for
the mass of the planet considering its contrast and assuming that the dust is
filling the Hill radius around the planet. The contrast of the object is given
by:
\begin{equation}
  \label{e:contrast}
    C=A\cdot{{\pi r_{H}^{2}}\over{4\pi a^2}}\cdot{{1}\over{k}}
\end{equation}
where A is the albedo, $r_{H}$ is the Hill radius, $a$ is the separation of
the object from the star and $k$ is a multiplicative factor depending on the
geometry of the system. Considering the inclination of the system, we found
that a reasonable value of $k$ is 1.38. The Hill radius is given by:
\begin{equation}
  \label{e:hill}
  r_{H}=a\cdot\sqrt[3]{{m_{p}}\over{3M_{\star}}}
\end{equation}
where $m_{p}$ is the mass of the planet and $M_{\star}$ is the stellar mass.
Joining equation~\ref{e:contrast} and equation~\ref{e:hill} we obtain an
expression for the planetary mass:
\begin{equation}
  \label{e:planmass}
  m_{p}=3M_{\star}\cdot\biggl({{4 C k}\over{A}}\biggr)^{{3}\over{2}}
\end{equation}
The value of the contrast is $7.27\times10^{-5}$ as obtained from a median of
the contrast values on each IFS spectral channel. We then assume for the star
a mass of 0.76~\MSun \citep{2018A&A...617L...2M} and an albedo ($A$) of 0.5.
We then obtain a value for the mass of
the planet of $5.2\times10^{-5}$~\MSun corresponding to $\sim$17.3~$M_{\oplus}$.
This is of course a lower limit valid in the case that the circumplanetary
material is filling the Hill radius. In a more evolved case, in which the
circumplanetary material is filling 1/3 of the Hill radius as proposed e.g. by
\citet{2009MNRAS.397..657A} the mass of the planetary objects would be of the
order of 90~$M_{\oplus}$. It this were true, in the PDS\,70 system we would have
planets at a different evolution stage. Indeed, while PDS\,70\,b and PDS\,70\,c
are accreting material probably from a circumplanetary disk, this object would
still be in its formation phase. This would probably recall what happened
in our solar system where the gaseous giant planets formed before the
inner planets.

\subsection{Stability of the system}
\label{s:stability}

\begin{figure}
\centering
\includegraphics[width=1\columnwidth]{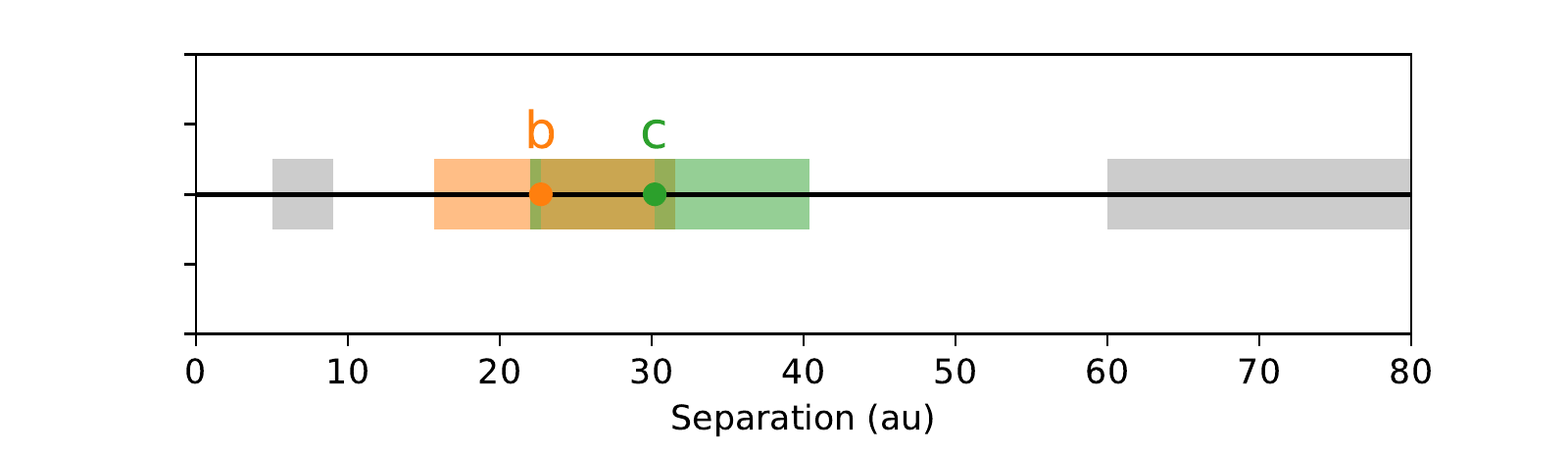}
\caption{Positions of the chaotic zones for PDS\,70\,b and PDS\,70\,c. The
  grey zones represent the disk positions.}
\label{f:zcline0}
\end{figure}

\begin{figure}
\centering
\includegraphics[width=1\columnwidth]{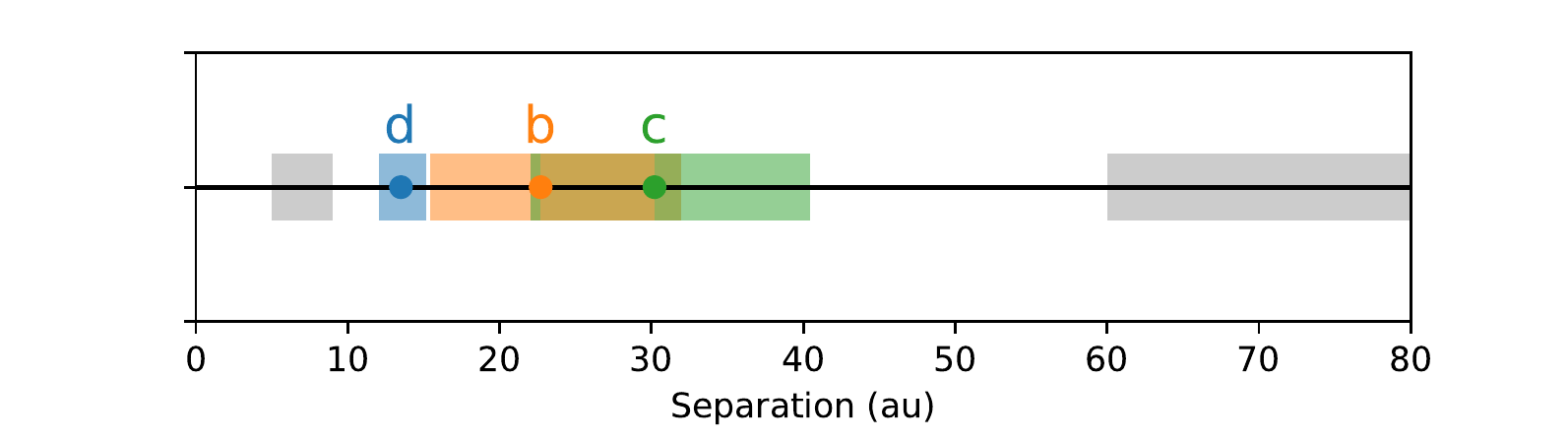}
\caption{Simular to Figure~\ref{f:zcline0} but including also the PLF (here
  tagged as d) considering it as a real companion.}
\label{f:zcline}
\end{figure}

To test the stability of the system, we performed a few N-body simulations
with the integrator SWIFT HJS \citep{beust2003}. This method is optimized to
check what happens in the disk cavity because it does not take into account
the gas component of the disk. The main assumption of this model is that there
is no major influence of the circumstellar disk on the planetary orbits. The
critical parameters that control the stability of the system are the semi-major
axes ratios and the eccentricities. To this aim we used the values listed in
Table~\ref{t:struct} for PDS\,70\,b and PDS\,70\,c. Given the current
observational constraints, the stability of the system is not guaranteed if the
masses are as large as obtained assuming an age of 5.4~Myr for the planets.
The companions are not sufficiently separated to avoid significant gravitational
interactions in a timescale of the order of few thousands of years that
is not compatible with the age of the system. The chaotic zones around each
one of them can be
computed from \cite{morrison2015}. The case of coplanar circular orbits at the
observed separations is represented in Figure~\ref{f:zcline0}. The separation
of PDS\,70\,c falls within the chaotic zone of PDS\,70\,b, indicating strong
dynamical perturbations.
If we assume that the PLF is actually a massive body with the mass determined
in Section~\ref{s:natd} the results on the stability do not change due to its
proposed low mass. As shown in Figure~\ref{f:zcline}, it is outside the chaotic
zones of its neighbor, but not by far, so that eccentricity fluctuations could
eventually destabilize it. Assuming circular orbits for all three companions,
to limit dynamical perturbations, still leads to a destabilization within a few
thousand years. Given the age of the system, this is not likely to be the case. 
The companion separations are compatible with a configuration of
mean-motion-resonances 1:2:3, and marginally, with a 1:2:4 configuration.
This system would be thus very similar to the 4-planet system HR 8799
\citep{marois2008}, who also appeared unstable and that is potentially in
resonant configuration \citep{gozdziewski2009} or stable at the condition that
the planets are coplanar or in mean-motion resonance
\citep{2016AJ....152...28K,2018AJ....156..192W}. A comprehensive study of the
dynamics of PDS\,70 would need to take into account both the mean-motion
resonances and the interactions with the disk, which could act as a stabilizing
factor by damping the eccentricities.
On the other hand, at first order, both the inner disk and outer disk edges
are compatible with the positions of the bodies. Furthermore, the disk is
susceptible to be affected by the gravitational influence of the PLF, if a real
object,  for the inner disk and PDS\,70\,c for the outer disk, through secular
dynamics and mean-motion resonances. This could cause local depletions in
some part of the disk or asymmetries. However, to study in more detail these
effects we would need hydrodynamical simulations of the disk evolution and
this is outside the scope of this paper. \par
We would like to stress that these results have been achieved assuming for
PDS\,70\,b and PDS\,70\,c masses obtained using the age system of 5.4~Myr. 
Anyhow, given that these two objects are still in-formation, their ages could
be younger and, as a consequence, also their masses would be lower changing
the results about the stability of the system. Indeed, stability simulations
performed with lower masses for PDS\,70\,b and PDS\,70\,c (2~\MJup), would
allow to reach the stability also when we change only the mass of one of
the two planets if the separation of PDS\,70\,c is 31~au or larger. The
orbital parameters of this object were determined thanks to only two
astrometric points so that further astrometric measures will be needed to
better constrain them. If we instead consider for these two objects the two
masses given in Table~\ref{t:struct}, we would need for PDS\,70\,c a separation
larger than 35-36~au or alternatively a separation for PDS\,70\,b smaller than
20~au, obviously just in the case that the PLF is not a companion. \par
The system stability could be enhanced if the two planets are in mean motion
resonance as found recently by \citet{2019arXiv190909476B} using two-dimensional
hydrodynamic simulations.

\section{Conclusions}
\label{s:conclusion}

In this work we presented the SPHERE observations aimed to verify the
presence of a second planetary companion around the star PDS\,70.
Our results confirm the presence of PDS\,70\,c both in J- and H-spectral bands
with IFS and in K band with IRDIS despite the flux is very low at shorter
wavelengths. We retrieved astrometric positions for this object in two
different epochs and this allowed us to support that this object is
actually gravitationally bound to the star. The extracted photometry
hints toward a very red spectrum that is very probably due to the presence
of dust in the planetary atmosphere and/or to the presence of a circumplanetary
disk as proposed through recent ALMA observations. We were able to obtain a
good fit with atmospheric models and our results clearly tend toward
a low gravity (3.0-3.5~dex) and low temperature ($\sim$900~K) object using
both the Exo-REM and the BT-Settl models. The mass evaluated through these
models is just in marginal agreement with that obtained through the
AMES-DUSTY evolutionary models, but we can however conclude that it should be
less than $\sim$5~\MJup. Finally, the probable presence of dust in the
atmosphere of this object and/or of a circum-planetary disk is also confirmed
by its position in the color-magnitude diagram that is even more extreme than
that of PDS\,70\,b. \par
In addition to the results on PDS\,70\,c, the analysis of archival and new
data allowed us to detect the presence of a possible third point source in this
planetary system (PLF). This object is located at small separation
($\sim$0.12\as) from the star and it is clearly detected only in the IFS data
while it is barely visible in the IRDIS data. In any
case, it was possible to detect it in four different epochs with a time range of
about three years allowing to exclude that it can simply be a feature due to
the speckle subtraction method. Also, these astrometric positions allowed us to
confirm that this object is actually gravitationally bound to the star. However,
the extracted spectrum for this object is blue hinting toward stellar light
reflected from dust. This result seems to demonstrate that this object is
actually a feature due to the inner disk. This is further confirmed by an
analysis of the non-coronagraphic PSF data from which we were able to find
hints of the presence of the inner disk at the separation of this feature.
We cannot, however, definitely exclude the possibility that
it is a separate object and in this case we considered two possible solutions.
The first one is that it is just a transient blob of dust while in the second
case it could be a forming planet completely embedded in a dust envelope. In
the first case we should assume that the generation of such blobs is frequent
through some not well defined mechanism. In the second case we would not be
able to see any light from the planetary photosphere and this would justify
the extracted spectrum. If this is the real nature of this object we can
evaluate its mass being less than 90~$M_{\oplus}$. It is not possible with the
present data to draw a definitive conclusion about the nature of this object.
To be able to obtain a definitive answer on this point, interferometric
data with instruments like e.g. GRAVITY \citep{2017A&A...602A..94G} of the
inner part of this system would be mandatory. \par
We also presented a possible solution for the general structure of the
planetary system assuming for the moment circular orbits for all the candidate
companions. Moreover, the simulations were performed both excluding the PLF and
both considering it. In both cases, the proposed configuration does not seem to
be stable on long periods if the masses are as large as obtained assuming that
the age of the companions is the same as for the star. In any case, as noted
above, lower masses of the companions due to a lower age of the forming planets
with respect to the system would allow us to obtain a stable configuration.
Clearly long period astrometric follow-up observations will be needed to
further constrain the orbits of the proposed objects and to draw more precise
conclusion about the structure and the stability of the PDS\,70 system. \par
From our data we can conclude that PDS\,70 hosts the second
multiple planetary system found with the direct imaging technique and the first
one at a very young age, where the planets are still accreting.


\begin{acknowledgements}
The authors thanks the anonymous referee for the constructive comments that
strongly helped to improve the quality of this work. 
This work has made use of the SPHERE Data Center, jointly operated by
OSUG/IPAG (Grenoble), PYTHEAS/LAM/CeSAM (Marseille), OCA/Lagrange (Nice) and
Observatoire de Paris/LESIA (Paris). \par
This work has made use of data from the European Space Agency (ESA) mission
{\it Gaia} (\url{https://www.cosmos.esa.int/gaia}), processed by
the {\it Gaia} Data Processing and Analysis Consortium (DPAC,
\url{https://www.cosmos.esa.int/web/gaia/dpac/consortium}). Funding for
the DPAC has been provided by national institutions, in particular the
institutions participating in the {\it Gaia} Multilateral Agreement. \par
This research has made use of the SIMBAD database, operated at CDS,
Strasbourg, France. \par
D.M., R.G., S.D., A.Z. acknowledge support from
the ``Progetti Premiali'' funding scheme of the Italian Ministry of Education,
University, and Research. A.Z. acknowledges support from the CONICYT + PAI/
Convocatoria nacional subvenci\'on a la instalaci\'on en la academia,
convocatoria 2017 + Folio PAI77170087. A.M. acknowledges the support of the
DFG priority program SPP 1992 "Exploring the Diversity of Extrasolar Planets"
(MU 4172/1-1). C.\,P. acknowledge financial support from Fondecyt (grant 3190691) and financial support from the ICM (Iniciativa Cient\'ifica Milenio) via the N\'ucleo Milenio de Formaci\'on Planetaria grant, from the Universidad de Valpara\'iso. T.H. acknowledges support from the European Research Council under the
Horizon 2020 Framework Program via the ERC Advanced Grant Origins 83 24 28.
\par
SPHERE is an instrument designed and built by a consortium consisting
of IPAG (Grenoble, France), MPIA (Heidelberg, Germany), LAM (Marseille,
France), LESIA (Paris, France), Laboratoire Lagrange (Nice, France),
INAF-Osservatorio di Padova (Italy), Observatoire de Gen\`eve (Switzerland),
ETH Zurich (Switzerland), NOVA (Netherlands), ONERA (France) and ASTRON
(Netherlands), in collaboration with ESO. SPHERE was funded by ESO, with
additional contributions from CNRS (France), MPIA (Germany), INAF (Italy),
FINES (Switzerland) and NOVA (Netherlands). SPHERE also received funding
from the European Commission Sixth and Seventh Framework Programmes as
part of the Optical Infrared Coordination Network for Astronomy (OPTICON)
under grant number RII3-Ct-2004-001566 for FP6 (2004-2008), grant number
226604 for FP7 (2009-2012) and grant number 312430 for FP7 (2013-2016).
\end{acknowledgements}

\bibliographystyle{aa}
\bibliography{pds70}

\end{document}